\documentclass{elsarticle}
\usepackage{geometry}
\usepackage{graphicx}
\usepackage{amssymb}
\usepackage{amsmath}
\usepackage[modulo]{lineno}
\usepackage{epstopdf}
\title{Parametric scaling of power exhaust in EU-DEMO alternative divertor simulations}
\author[VTT]{A.E. Järvinen\corref{cor1}}
\ead{aaro.jarvinen@vtt.fi}
\author[VTT]{L. Aho-Mantila}
\author[IPP]{T. Lunt}
\author[TORINO]{F. Subba}
\author[TUSCIA]{G. Rubino}
\author[CCFE]{L. Xiang}

\cortext[cor1]{Corresponding author: aaro.jarvinen@vtt.fi}
\address[VTT]{VTT Technical Research Centre of Finland, FI-02044 VTT, Finland}
\address[IPP]{Max-Planck Institut für Plasmaphysik, D-85748 Garching, Germany}
\address[TORINO]{NEMO Group, Politecnico di Torino, Corso Duca defli Abruzzi 24, 10129 Torino, Italy}
\address[TUSCIA]{University of Tuscia, Largo dell’Università snc, Viterbo, 01100, Italy}
\address[CCFE]{CCFE – UKAEA, Culham Science Centre, Abingdon, OX14 3DB, UK}

\begin{document}
\begin{abstract}
    Investigations of parametric scaling of power exhaust in the alternative divertor configuration (ADC) SOLPS-ITER simulation database of the EU-DEMO are conducted and compared to predictions based on the Lengyel model. The Lengyel model overpredicts the necessary argon concentrations for LFS divertor detachment by about a factor of 5 – 10 relative to the SOLPS-ITER simulations. Therefore, while the Lengyel model predicts that plasmas with accetable divertor heat loads in EU-DEMO would exceed the tolerable upstream impurity concentrations by a large margin, there are several SOLPS-ITER solutions within an acceptable operational space. The SOLPS-ITER simulations indicate that, unlike assumed by the standard Lengyel model, there are significant heat dissipation mechanisms other than argon radiation, such as cross-field transport, that reduce the role of argon radiation by a factor of 2 to 3. Furthermore, the Lengyel model assumes that the radiation front is powered by parallel heat conduction only, which tends to lead to a narrow radiation front as the radiative efficiency increases strongly with reducing thermal conductivity. As a result, the radiative volume and total impurity radiation are suppressed for a given impurity concentration. However, the SOLPS-ITER simulations indicate that other mechanisms, such as cross-field transport, can compete with parallel heat conduction within the radiative front and increase the radiative volume. Due to these findings, usage of the standard Lengyel model for analyzing scaling between divertor conditions and configurations for devices such as EU-DEMO is strongly discouraged. 
    
\end{abstract}


\maketitle

\section{Introduction}
Power exhaust is one of the main challenges faced by reactor-scale fusion devices. To address the risk that a conventional divertor, as pursued in ITER, may not necessarily extrapolate to a DEMO reactor, the EUROfusion consortium has been systematically investigating the benefits and challenges of alternative divertor configurations (ADC) \cite{MilitelloNME2021, ReimerdesNF2020, XiangNF2021, AhoMantilaNME2021, SubbaNME2017, SubbaPPCF2018, TheilerPSI2022, ThorntonPSI2022}. This effort has produced a large database of ADC simulations for the EU-DEMO, conducted with the SOLPS-ITER code package \cite{XiangNF2021, AhoMantilaNME2021, SubbaNME2017, SubbaPPCF2018, WiesenJNM2015}. While SOLPS-ITER simulations provide state-of-the-art predictions for the exhaust performances of the various divertor configurations, the predictions are sufficiently non-linear that is typically challenging to infer unambiguous causalities, dependencies, and conclusions about the obtained solutions. To address the need for a tractable and easily inferable model, several reduced models have been developed for predicting and analyzing the scrape-off layer (SOL) and divertor performance \cite{Lengyel, HutchinsonNF1994, PostPoP1995, MatthewsJNM1997, KallenbachPPCF2016, LipschultzNF2016, ReinkeNF2017, GoldstonPPCF2017}. Most of these follow the Lengyel model \cite{Lengyel}, which uses simplifying assumptions about SOL transport to relate the SOL impurity density, upstream electron density, and the upstream SOL heat flux to the dissipated power and onset of detached conditions. However, while the Lengyel model does provide a convenient tool to estimate the scaling of the power dissipation by impurity radiation, it neglects a self-consistent treatment for several important physical processes, such as cross-field transport, cooling due to interaction with the neutral population in the divertor, convective energy transport, or changes of flux expansion within the divertor leg. Many of these physical processes are the key features that separate ADCs from conventional configurations, and, therefore, the simple Lengyel model is not expected to appropriately capture the key benefits of ADCs. Despite this, in a previous study \cite{XiangNF2021}, the Lengyel model was observed to qualitatively capture the impact of increased SOL connection length between the Super-X (SX) and conventional single-null (SN) configurations on the required argon concentration in the SOL to achieve detached conditions. Similarly, in the study by Moulton et al. for the ITER SOLPS-4.3 database \cite{MoultonNF2021}, the relative interdependencies between the electron density, parallel heat flux, and impurity concentration in the outer divertor predicted by Lengyel model were remarkably accurate \cite{MoultonNF2021}. However, the absolute detachment threshold impurity concentration was overpredicted by a factor of 4.3 by the Lengyel model relative to that predicted by SOLPS-4.3. 

In this study, the parametric scaling of power exhaust in these EU-DEMO simulations are investigated by comparing the predicted performance of the SX and X-divertor (XD) configurations to the conventional SN configurations. 
The investigated SOLPS-ITER simulations were conducted with reduced physics models designed to facilitate throughput for reactor scoping studies \cite{CosterCPP2016}. The simplifications include fluid model for neutrals, omission of cross-field drifts, and bundling of the argon impurity representation to three fluid species: neutral argon, all ionization levels that are not fully stripped, and fully stripped argon. Further information about the setup of the SOLPS-ITER simulations can be found in the previous publications \cite{XiangNF2021, AhoMantilaNME2021, SubbaNME2017, SubbaPPCF2018}. 

\section{Methodology}


\begin{figure*} [!htb]
    \centering
    \includegraphics[width=0.9\textwidth]{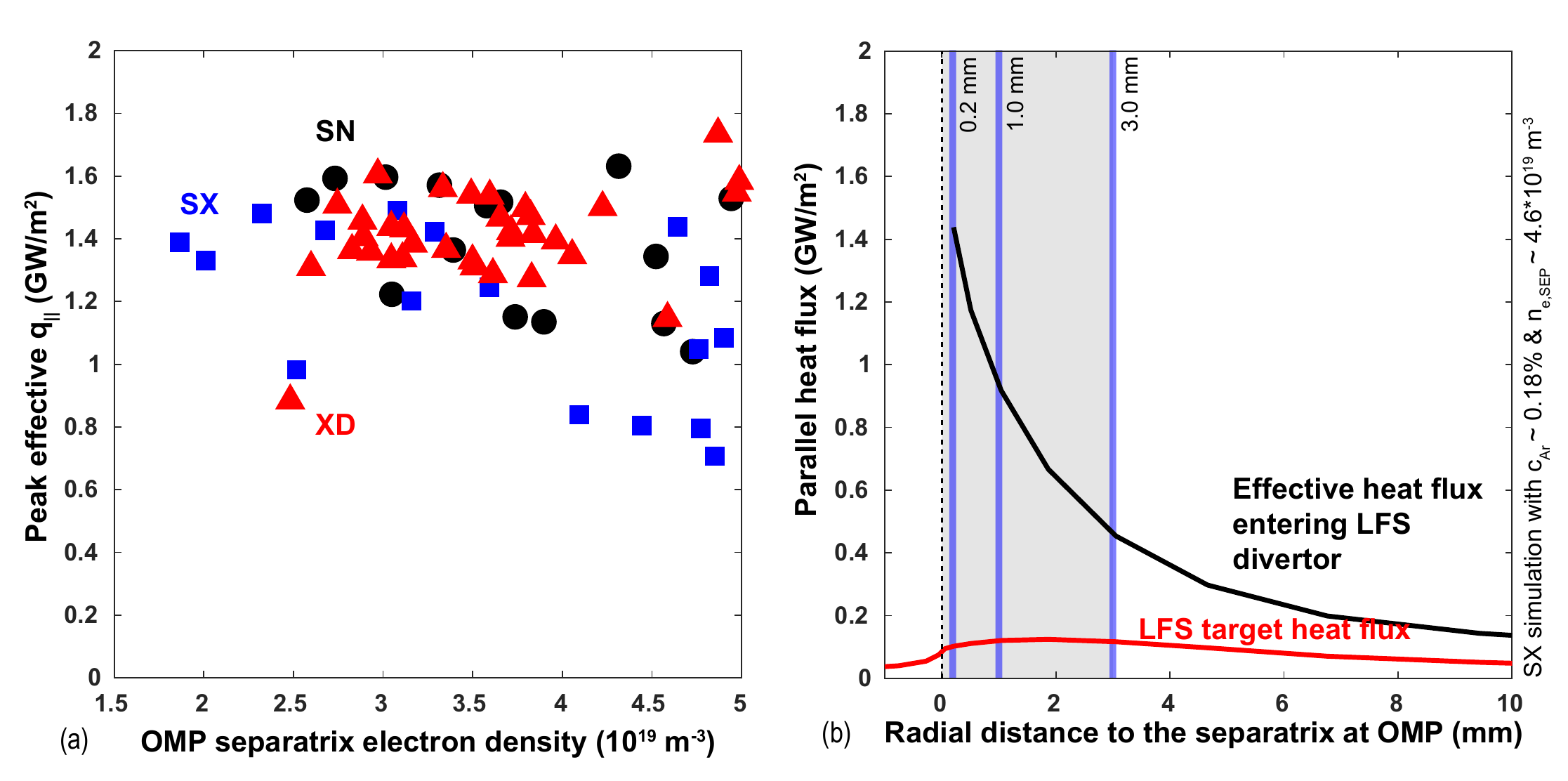}
    \caption{(a) Peak of the effective $q_\parallel$ at the LFS divertor entrance as a function of OMP separatrix electron density. The symbols and colors are the same as in Figure 1. (b) Example of the effective parallel heat flux profile at the LFS divertor entrance (black line) and at the LFS target (red line) in an SX simulation (MDS+ identifier 170182). The radial locations of 0.2 mm, 1.0 mm, and 3.0 mm from the separatrix at the OMP are illustrated with narrow blue bars. The gray background illustrates the area for which the dissipation integrals are calculated in Figure \ref{fig_diss_integrals}}
    \label{fig_heat_flux}
\end{figure*}

The analysis in this manuscript is focused on the low field side (LFS) SOL and divertor. This is the divertor that usually receives the greatest steady-state heat fluxes, and due to omission of cross-field drifts, the divertor asymmetries between the high field side (HFS) and LFS are expected to be underestimated in the simulations investigated in this study. However, this does not mean that the HFS divertor would not be important, and in some cases when the LFS divertor dissipation is enhanced through ADCs, the HFS divertor might become the divertor that limits the extent of the acceptable operational window. 

The investigated database is limited to solutions that are within the acceptable margin in terms of peak divertor plate heat fluxes below 5 MW/m$^2$ and peak electron temperature at the divertor plates below 5 eV. Furthermore, only solutions with separatrix electron density, $n_\text{e,SEP}$, below $5 \times 10^{19}$ m$^{-3}$ and separatrix argon concentration, $c_\text{Ar,SEP}$, below 2\% are included in the analysis. These are somewhat wider ranges than documented as the acceptable operational window in the report by Xiang et al. \cite{XiangNF2021}: $n_\text{e,SEP}$ below $4.2\times 10^{19}$ m$^{-3}$, which represents about 60\% of the Greenwald density limit, and $c_\text{Ar,SEP}$ below 1\%. The rationale is to include solutions somewhat outside the operational window into the analysis, while still limiting the database sufficiently to focus the analysis to solutions that would be considered to be within the operational space. Furthermore, any plasma solutions with power crossing the separatrix, $P_\text{SEP}$, below 100 MW are not included in the analyzed dataset. The input power in the simulations is 150 MW and $P_\text{SEP}$ below 100 MW indicates significant radiation inside the separatrix, which raises uncertainty about the high confinement mode (H-mode) relevance of such plasma solutions. While it is possible that these plasmas would eventually approach an X-point radiator operational regime, which can retain H-mode like confined plasma properties \cite[and references therein]{BernertNF2021}, the detailed study of these plasma solutions is not included in this work. Considering the other constraints on the investigated operational space, this $P_\text{SEP}$ constrain only filters out the XD simulations with argon seeding rate exceeding $3.5\times 10^{21}$ s$^{-1}$. Furthermore, any cases that employ significantly stronger argon puff than is needed to reach detached conditions are filtered out. This results in removing all but the lowest density XD solution at argon seeding rate $3.5\times 10^{21}$ s$^{-1}$ from the analyzed dataset. 

$n_\text{e,SEP}$ is evaluated at the separatrix at the outer mid-plane (OMP).  $c_\text{Ar}$ is taken as an average between the X-point and the OMP between the separatrix and the flux surface 1 mm radially away from the separatrix at the OMP. These definitions are somewhat different than those used by Moulton et al. \cite{MoultonNF2021}, where the average impurity concentration in the divertor leg was used and the electron density was taken at the divertor entrance of the analyzed SOL ring. While the latter approach provides a cleaner comparison to the Lengyel model in the divertor leg, in this study for EU-DEMO the more experimentally motivated upstream conditions for $n_\text{e,SEP}$ and $c_\text{Ar}$ were selected. This selection is also motivated by the need to establish a common set of upstream boundary conditions for analyzing different ADCs as well as for conditions ranging all the way to very strong detachment with temperatures collapsed to low values at the divertor entrance.

The main driving parameter for the SOL plasma is the heat flux entering the SOL upstream, $q_{\parallel\text{,u}}$. Defining $q_{\parallel\text{,u}}$ in a systematic fashion that allows unambiguous comparison between SOLPS-ITER calculations and a simple Lengyel model approach can be challenging. In the work by Moulton et al. \cite{MoultonNF2021}, $q_{\parallel\text{,u}}$ was taken at the LFS divertor entrance of SOL ring number three from the separatrix. For analysis of a database of ITER simulations, this is an appropriate choice, since the divertor plasma solutions are primarily partially detached, and most of the dissipation occurs below the divertor entrance. For EU-DEMO database, this assumption does not hold, and $q_{\parallel\text{,u}}$ at the divertor entrance is expected vary significantly between solutions within the operational window, due to dissipation above the X-point. To take this into account, an effective $q_{\parallel\text{,u,eff}}$ is calculated by including the power dissipated in the SOL between the X-point and the OMP. This can be obtained by integrating the energy current entering or leaving the flux tube radially between the OMP and the X-point and poloidally at the OMP. Furthermore, the mapping from the total poloidal effective energy current for a given flux tube to the parallel heat flux is conducted by evaluating the flux tube poloidal area times the poloidal magnetic field component at the OMP. This avoids discretization errors near the X-point that can be quite significant as the poloidal magnetic field asymptotically approaches zero, while the magnetic flux within a flux tube should stay constant. The toroidal field magnitude is taken at the entrance to the divertor, as $q_{\parallel\text{,u}}$ is linearly proportional to the magnetic field strength. As a result, even if the total energy current would stay constant in a flux tube, the $q_{\parallel\text{,u}}$ would change according to any changes of $B_\text{T}$. 

A further challenge in selecting the appropriate $q_{\parallel\text{,u}}$ is the radial location for evaluating $q_{\parallel\text{,u}}$. The highest $q_{\parallel\text{,u}}$ occurs typically near the separatrix, but these flux tubes also experience significant cross-field heat transport losses in the divertor, such that usually they are not the limiting flux tubes for plate power fluxes. Due to this reason, the ring number three from the separatrix was chosen in the work by Moulton et al. \cite{MoultonNF2021}. However, one of the key limitations of the simple models is exactly the fact that cross-field heat losses are often not included systematically, while usually the maximum upstream heat flux is chosen as the challenge metric in the simple analysis. Therefore, in this work, the maximum effective $q_{\parallel\text{,u}}$ at the divertor entrance is taken as the challenge metric and the discrepancy caused by cross-field transport is analyzed as a shortcoming of the simple model approach. With this approach, $q_{\parallel\text{,u,eff}}$ is about 1.0 – 1.6 GW/m$^2$ (Fig. \ref{fig_heat_flux}). At low density and impurity injection rates, the value is close to the upper envelope around 1.5 - 1.6 GW/m$^2$. With increasing $n_\text{e,SEP}$, the heat flux is reduced as cross-field heat diffusion flux is linearly proportional to density ($q_\text{r, cond} = n\chi\partial{T}/\partial{r}$). With increasing $c_\text{Ar}$, $P_\text{SEP}$ is reduced as argon is an efficient radiator in temperatures relevant for plasmas inside the separatrix, reducing the $q_{\parallel\text{,u}}$ as well. When the divertor collapses to strongly detached conditions, with the radiative front at the X-point level, the heat flux at the flux tube next to the separatrix increases again to the level close to 1.6 GW/m$^2$ in many of the cases due to the strong temperature gradients accelerating heat flow to the radiative front. This effectively reduces the radial decay length of the heat flux. Due to these multiple competing processes, clear trends are difficult to observe in Figure 1a that encompasses several fuel and impurity gas injection levels. Figure \ref{fig_heat_flux} shows an example $q_{\parallel\text{,u,eff}}$ profile entering the LFS divertor and the $q_{\parallel}$ at the LFS target for an example SX simulation. 

 \begin{table}[]
    \centering
    \caption{Connection lengths between the OMP and the LFS target at about 0.2 mm, 1.0 mm, and 3.0 mm from the separatrix at the OMP for the analyzed configurations. These flux lines are illustrated relative to the heat flux profile in Figure \ref{fig_heat_flux}b}
    \begin{tabular}{|c|c|c|c|}
    \hline
    $R_\text{OMP} - R_\text{SEP}$ (mm) &	SN (m) & SX	(m) & XD (m) \\
    \hline
    0.2 & 152 &	253 & 291 \\
    1.0 & 118 &	201 & 235 \\
    3.0 & 102 &	180 & 205 \\
    \hline
    \end{tabular}
    \label{tab1}
\end{table}

A further choice that must be made in the comparison is the selection of connection length, $L_\text{CONN}$. As the poloidal field approaches zero at the X-point, the connection length is significantly longer near the separatrix than radially away from the separatrix (Table \ref{tab1}). Between 0.2 mm and 3.0 mm from the separatrix at OMP, $L_\text{CONN}$ is reduced by about 30 \% (Table 1). However, integrating a Lengyel model type equation systems, one is free to continue the integration up to an arbitrary $L_\text{CONN}$, providing a natural approach to address the impact of $L_\text{CONN}$ uncertainty to the solution. 

\section{Comparison to the Lengyel model}
\begin{figure} [!htb]
    \centering
    \includegraphics[width=0.4\textwidth]{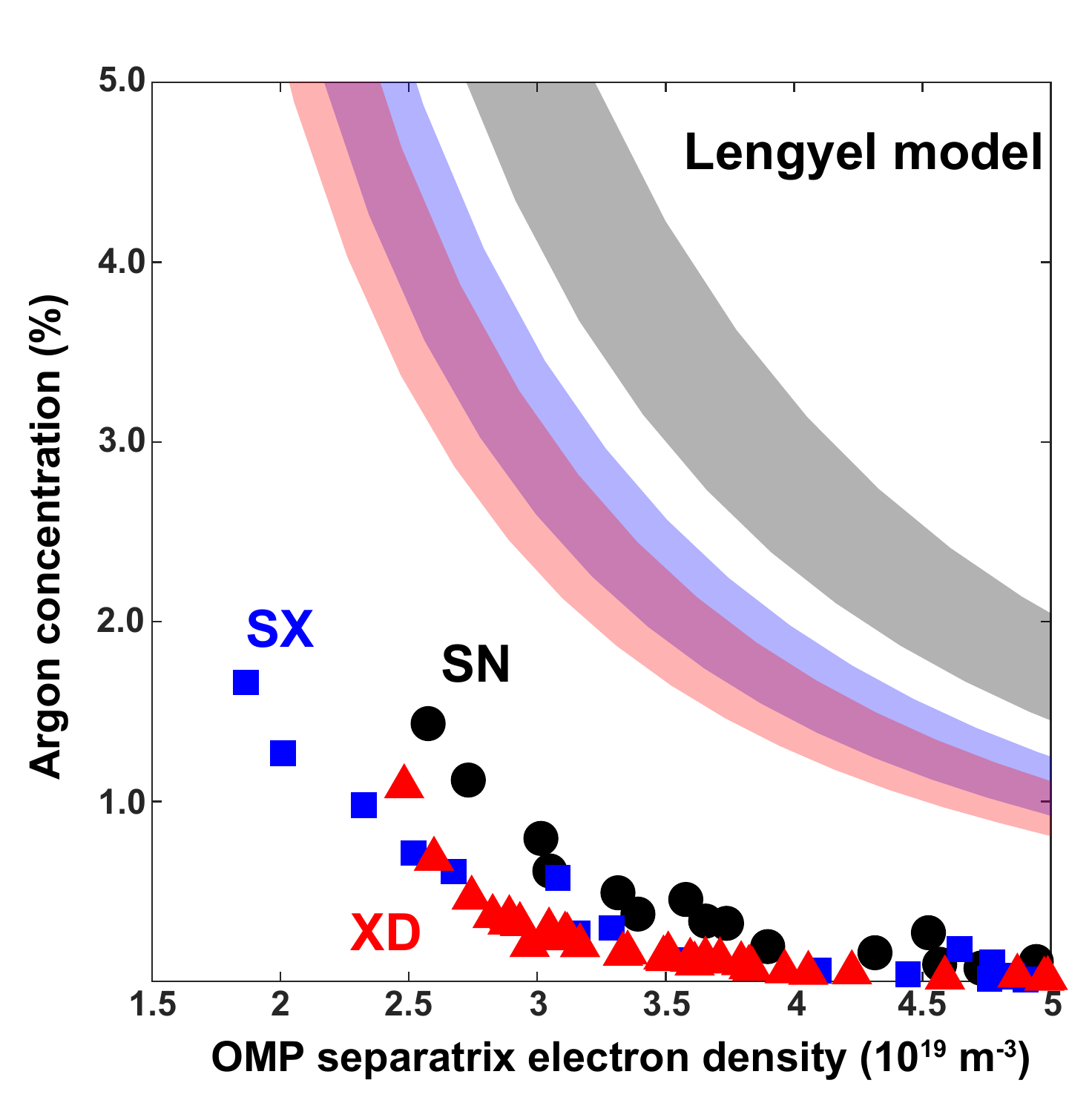}
    \caption{Comparison of Lengyel model and SOLPS-ITER predictions for the upstream argon concentration required for LSF divertor detachment as a function of OMP separatrix electron density. The width of the Lengyel model bars represent the upper and lower ranges of the connection lengths between 0.2 and 3.0 mm from the separatrix at the OMP (Table \ref{tab1}).}
    \label{fig_Lengyel}
\end{figure}

The Lengyel model overpredicts the necessary argon concentration for LFS divertor detachment about a factor of 5 – 10 relative to the SOLPS-ITER simulations (Fig. \ref{fig_Lengyel}). This is consistent with the previous study by Moulton et al. for the ITER SOLPS4.3 database, showing a factor of 4.3 overprediction \cite{MoultonNF2021}. Clearly the Lengyel approach would provide a very pessimistic projection to reactor-scale facilities, such as EU-DEMO. Basically, with the given parameters no acceptable solutions would exist within the operational window. This is consistent with projections provided by the recent power exhaust and detachment scaling studies by Goldston and Reinke that would project unacceptably high impurity concentrations for an EU-DEMO type device \cite{ReinkeNF2017, GoldstonPPCF2017}. However, SOLPS-ITER simulations for EU-DEMO, fortunately, indicate that significantly lower impurity concentrations might be sufficient to reach LFS divertor detachment (Fig. \ref{fig_Lengyel}).  

The upstream heat flux in these Lengyel model calculations is set to 1.2 GW/m$^2$, representing roughly the lower envelope for bulk of the scatter in the analyzed dataset (Fig. \ref{fig_heat_flux}a). Heat flux limit was applied in the heat conduction calculations with the limit set to 0.2, similar to the SOLPS-ITER simulations. Target $T_e$ was set to 2.0 eV.  Target $n_e$ is scanned to scan the density in the SOL. A factor of 2 static pressure drop is assumed right in front of the target due to the acceleration to sonic flow at the sheath edge. Sheath heat transmission coefficient was set to 7.0. The thermal conductivity without the temperature dependence, $\kappa_0$, was set to 2600 Wm$^{-1}$eV$^{-7/2}$. The impact of $Z_\text{eff}$ on thermal conductivity was neglected. Argon cooling is calculated based on ADAS \cite{ADAS} data assuming $n_\text{e}\tau \sim 10^{16}$ m$^{-3}$s. This assumption is actually enhancing argon radiation rates relative to the SOLPS-ITER simulations, which due to the chosen argon bundling scheme show cooling rates close to the equilibrium values. However, $n_\text{e}\tau \sim 10^{16}$ m$^{-3}$s is a relatively standard assumption in Lengyel model type calculations and is, therefore, chosen here as a representative value for what to expect in a Lengyel model approach. Assuming coronal equilibrium rates would increase the threshold impurity concentrations in the Lengyel model calculations by about 50\%. The width of the Lengyel model bars represent the envelope of solutions between $L_\text{CONN}$ at 0.2 and 3.0 mm from the separatrix at the OMP (Table \ref{tab1}).


In the standard Lengyel model, the only parameter that changes between the divertor configurations is $L_\text{CONN}$. Based on $L_\text{CONN}$ changes between the configurations, the Lengyel model would predict the SX and XD configurations to reach detachment at argon concentrations about 55 – 60\% of those of the SN configuration (Fig. \ref{fig_Lengyel}). This seems also qualitatively consistent with the SOLPS-ITER simulations, and it would be tempting to conclude that the Lengyel model is capturing the key dependency. However, as will be analyzed in Section 4, given all the other significant physics and dissipation mechanisms, not included in the Lengyel model, such a conclusion would be unlikely to be actually true.   

According to the standard Lengyel model, one would expect argon concentration to scale as $n_\text{e,SEP}^{-2}$. However, based on the SOLPS-ITER simulations in the analyzed dataset, the actual scaling is significantly closer to $n_\text{e,SEP}^{-4}$. This was also observed by Xiang et al. \cite{XiangNF2021}. In recent experimental studies with JET and AUG by Henderson et al., a scaling of about $n_\text{e,SEP}^{-2.7}$ was found which is also stronger than predicted by standard Lengyel model, but weaker than indicated by the SOLPS-ITER simulations for the EU-DEMO database \cite{HendersonIAEA2021}. These observations indicate that the $n_\text{e,SEP}$ scaling of the standard Lengyel model also does not provide an accurate projections between various conditions and facility scales. 

Due to the limited extent of this manuscript, discussion about power scaling is not included. In the study by Xiang et al. \cite{XiangNF2021}, SOLPS-ITER simulations for the SX configuration with the input power doubled to 300 MW were investigated, indicating roughly a linear scaling of $c_\text{Ar}$ with heating power, which seems consistent with Lengyel model type predictions: $c_\text{Ar,SEP} \propto q_{\parallel\text{,u}}^{8/7}$ \cite{XiangNF2021, ReinkeNF2017}. 

\section{Analysis}

\begin{figure} [!htb]
    \centering
   \includegraphics[width=0.4\textwidth]{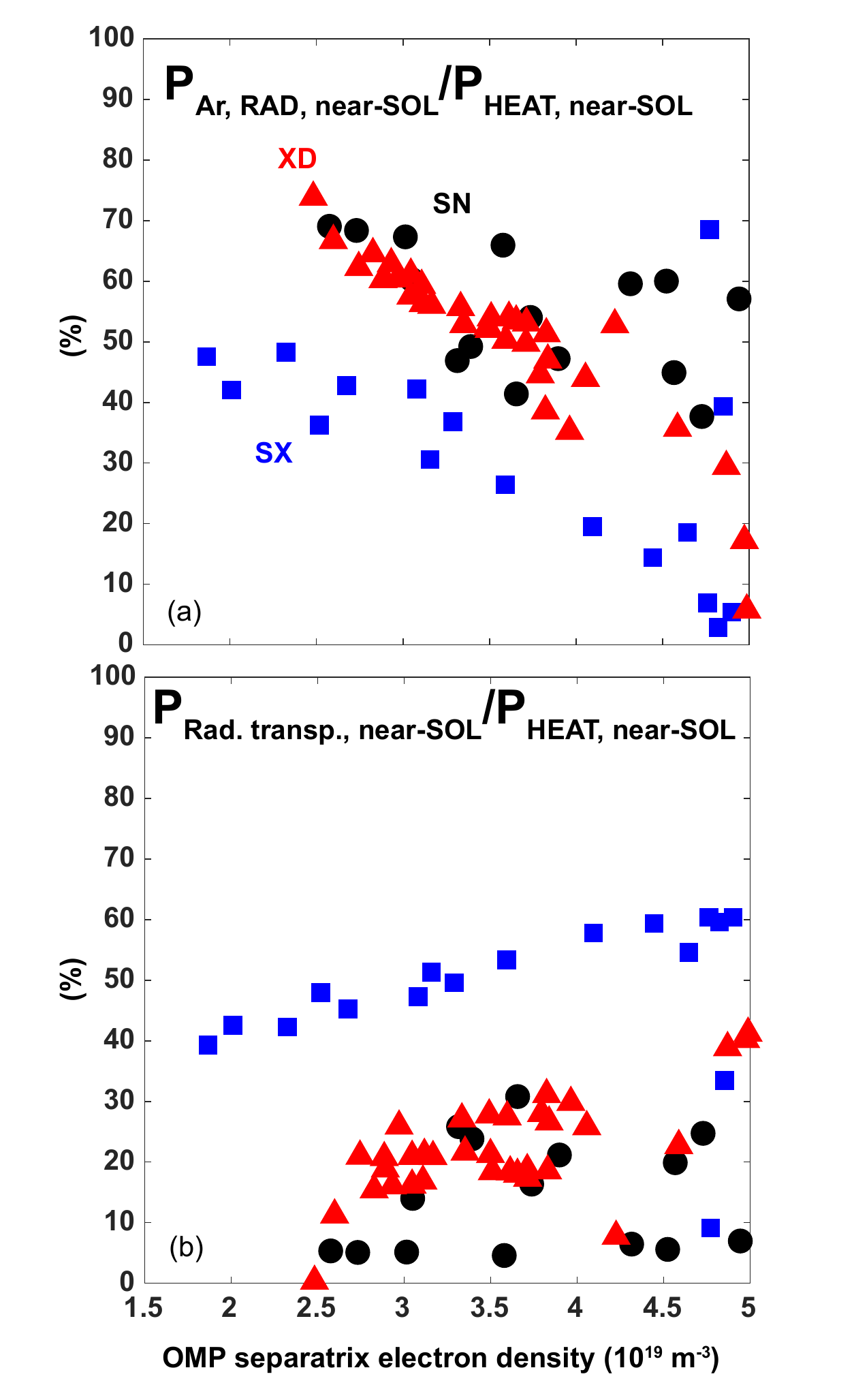}
    \caption{(a) Total electron cooling power due to argon in the LFS SOL below the OMP within 0 to 3 mm from the separatrix at the OMP. (b) Radial transport losses of heat in the LFS divertor from the LFS SOL region between 0 and 3 mm from the separatrix at the OMP. 3 mm represents roughly the radial heat flux width in these simulations (Figs. \ref{fig_heat_flux}). The values are normalized by the effective heat entering the near-SOL region upstream.}
    \label{fig_diss_integrals}
\end{figure}

 While the Lengyel model approach assumes that 100\% of the dissipation is provided by impurity radiation, the SOLPS-ITER database shows that in the near-SOL region in the LFS for most of the SN and XD simulations the fraction is between about 40 - 65\% (Fig. \ref{fig_diss_integrals}a). Near-SOL is defined as the area between the separatrix and 3 mm into the SOL from the separatrix measured at the OMP. This area roughly represents the distance of one heat flux width in these simulations (Fig. \ref{fig_heat_flux}b). Integrating through this region is expected to capture most of the dissipation processes required for the peak heat flux mitigation in the divertor as well as to sufficiently average over any details of cross-field transport between the high power density flux tubes. 
 In the SX configuration, the role of argon radiation is only between 20 - 40\% for most of the simulations (Fig. \ref{fig_diss_integrals}a). These observations indicate that the detachment threshold $c_\text{Ar}$ could in SN and XD cases already be reduced by about a factor of 2 from the Lengyel predictions, assuming linear or nearly linear scaling of exhaust power with $c_\text{Ar}$, and as much as a factor of 3 - 4 in SX. 
 
 The most significant dissipation process competing with argon radiation in the LFS near-SOL is cross-field transport to private-flux region (PFR) and to far SOL (Fig. \ref{fig_diss_integrals}b). For the SX configuration, this term is about a factor of 2 larger than for the SN and XD configurations. It seems that as the SX configuration tends to maintain relatively long connection lengths in the divertor at plasma temperatures above 20 eV, the radial heat diffusion model ($q_\text{r, cond} = n\chi\partial{T}/\partial{r}$) is able to dissipate the peak heat fluxes in the near-SOL. For most of the SX cases in this study, the connection length for the region above 20 eV temperature in the flux tube 2 mm from the separatrix at OMP is very close to 110 m until the grid cell right below the X-point. For the SN cases, this distance is between 0 and 30 m and for the XD cases, mostly between 40 to 80 m. Furthermore, due to the large major radius of the LFS divertor leg in SX, the total surface area between the divertor leg common SOL and the PFR is significantly larger, 280 m$^2$, than in the SN, 80 m$^2$, or XD, 130 m$^2$, configurations, further enhancing the radial transport loss of heat. A further reduction of the critical argon radiation level for detachment in SX is provided by the 30\% reduction of $B_\text{T}$ between the X-point and the target, which reduces the parallel heat flux and is one of the primary design features of this configuration.
 
 Obviously, any benefit that SX may obtain from cross-field transport phenomena depends on the assumption on radial heat diffusivity, which is simulated by user prescribed diffusion and conduction coefficients in SOLPS-ITER. However, as long as the heat transport flux is proportional to temperature or its gradient, one would expect radial heat transport to be enhanced by the type of divertor plasma solutions observed in the SX configuration. The scaling of $\chi$ or a potential convective process with the divertor conditions and geometry remains, of course, uncertain here and may impact this result. 
 
 Even though argon radiation does not provide 100\% of the dissipation, the total amount of argon dissipation observed in the SOLPS-ITER simulations still exceed the the amount expected in Lengyel model type calculations at those concentrations. Clearly, SOLPS-ITER simulations are indicating more radiative dissipation for a given $c_\text{Ar}$ at given upstream conditions than expected based on Lengyel model calculations. These differences can be driven either by $c_\text{Ar}$ varying poloidally, invalidating the poloidally constant assumption in the Lengyel model, or background plasma evolving poloidally differently than what would be expected based on conservation of static pressure and heat transported solely via electron heat conduction. Non-coronal effects are not expected to cause the difference here due to the bundling scheme used in these simulations. 
 
 Divertor enrichment of argon,  $c_\text{Ar, divertor}/c_\text{Ar, upstream}$, was investigated to address the question on poloidal distribution of $c_\text{Ar}$. $c_\text{Ar, divertor}$ is defined as the integral value within the divertor for the same flux tubes as the definition of $c_\text{Ar, upstream}$ given in Section 2. Within the investigated database, the mean value for the SN simulations is 1.14, for the SX cases 1.24, and for the XD cases 1.17. If restricting the analysis only to simulations with $n_\text{e, SEP}$ between 3.0 and 4.0$\times10^{19}$ m$^{-3}$, the ratios are 1.17 for SN, 0.77 for SX, and 1.21 for XD. while these numbers show that there is likely some configuration to configuration variation in terms of divertor enrichment of radiative impurities, the values are not sufficiently above 1 to explain the difference to the predictions provided by the Lengyel model. Therefore, it is likely that variations of the background plasma profiles deviate sufficiently from the simple analytical rules used in the Lengyel model to explain the difference in the radiative efficiency. 
 
 A key assumption in a Lengyel model type analysis is that heat is transported solely via parallel heat conduction, which scales as $T_e^{5/2}$. As a result, the model tends to generate a narrow radiation front, as the cooling rates increase strongly at the same time with reducing thermal conductivity as plasma temperatures are reduced. The radiation front is essentially power starved within a spatially narrow region, restricting the radiative volume and total radiation for a given $c_\text{Ar}$. In this situation, any transport mechanism that can keep on powering the high radiation zone can significantly enhance the total radiative volume and power \cite{JaervinenCPP2020}. Parallel convection, cross-field drifts, and cross-field conductive and convective heat transport can all compete with parallel heat conduction when $T_e$ is reduced below 20 eV. While the SOLPS-ITER simulations investigated in this study do not include cross-field drifts, the other two transport channels can provide heat transport to the radiative zone. 
 
 To diagnose this, the heat transport contributions due to ions and electrons both radially and poloidally to each grid cell in the near-SOL at $T_e$ below 20 eV are calculated and the argon radiation weighed mean is analyzed. In the SN cases, on average 79\% of the argon radiation in the LFS near-SOL takes place in plasma temperatures below 20 eV. In these temperatures, the poloidal electron heat transport channel provides only about 68\% of the total transport to the grid cells and estimating the conductive part of that would contribute to 52\%. For the SX configurations these numbers are 63\%, 74\%, and 51\%, and for the XD configurations 81\%, 59\%, and 40\%. Basically the simulations strongly indicate that the basic assumption of the Lengyel model that the peak radiative power zone is primarily heated through parallel electron heat conduction is inaccurate. This is due to the parallel convection and cross-field transport in the simulations, which become significant as the impurity radiation zone overlaps with the spatial region where recycling and fuel re-ionization takes place. While it would be tempting to assume that the two zones can be spatially separated, due to the $T_e$ difference between the primary fuel re-ionization ($T_e < 5 - 10$ eV) and peak impurity radiation zones ($T_e > 10$ eV), actually the simulations indicate that the there is significant mixing between the two zones, such that parallel convection and cross-field heat transport between flux surfaces can transport heat to high radiation, low $T_e$ zones and enhance the total radiation for a given $c_\text{Ar}$. 

\section{Conclusions}

Investigations of parametric scaling of power exhaust in the ADC SOLPS-ITER simulation database of the EU-DEMO are conducted and compared to predictions based on the Lengyel model. The Lengyel model overpredicts $c_\text{Ar}$ for LFS divertor detachment by about a factor of 5 – 10 relative to the SOLPS-ITER simulations. The SOLPS-ITER simulations indicate that, there are significant heat dissipation mechanisms other than argon radiation in the LFS near-SOL, such as cross-field transport, that reduce the role of argon radiation by a factor of 2 to 3. Furthermore, the Lengyel model assumes that the radiation front is powered by parallel heat conduction only, which tends suppress the radiative volume and total impurity radiation for a given $c_\text{Ar}$. However, the SOLPS-ITER simulations indicate that other mechanisms, such as cross-field transport and parallel convection, can compete with parallel heat conduction within the radiative front in the divertor, enhancing the total radiative power for a given $c_\text{Ar}$. Due to these findings, usage of the standard Lengyel model for analyzing scaling between divertor conditions and configurations for devices such as EU-DEMO is strongly discouraged. 

\section*{Acknowledgements}
This work has been carried out within the framework of the EUROfusion Consortium, funded by the European Union via the Euratom Research and Training Programme (Grant Agreement No 101052200 — EUROfusion). Views and opinions expressed are however those of the author(s) only and do not necessarily reflect those of the European Union or the European Commission. Neither the European Union nor the European Commission can be held responsible for them.

\end{document}